\documentstyle[preprint,aps,tighten]{revtex}
\begin{document}
\draft

\title{Chiral quark model with relativistic kinematics}

\author{H. Garcilazo\footnote{Permanent address: 
Escuela Superior de F\' \i sica y Matem\'aticas, 
Instituto Polit\'ecnico Nacional, Edificio 9,
07738 M\'exico Distrito Federal, Mexico}
and A. Valcarce} 

\address{Grupo de F\'\i sica Nuclear, 
Universidad de Salamanca, \\
E-37008 Salamanca, Spain}

\maketitle

\begin{abstract}

The non-strange baryon spectrum 
is studied within a three-body model that incorporates 
relativistic kinematics. We found that the combined
effect of relativistic kinematics together with the
pion exchange between quarks is able to
reverse the order of the first positive- 
and negative-parity nucleon excited states as 
observed experimentally. Including the chiral partner of the pion
(the $\sigma$ meson) leads to an overall
good description of the spectrum.

\end{abstract}


\narrowtext
\newpage

The effect of relativity in the bound state problem of three
quarks is expected not to be negligible since the 
constituent mass of the light quarks ($u$ and $d$)
is of the same size or even smaller than the excitations 
observed in the baryon spectrum. Therefore, the first step towards
a fully relativistic theory of the three-quark problem should be
to consider relativistic kinematics which in turn requires the
solution of the Faddeev equations for the three-body
problem in momentum space. 

The method to solve the Faddeev equations in momentum space 
when the interaction between quarks includes a confining 
potential has been recently described 
\cite{GAR1}. In particular,
the problem of three quarks interacting 
via a one-gluon-exchange (OGE) potential plus 
a linear confining interaction has been studied using both the 
nonrelativistic Faddeev equations and a version of 
these equations with fully relativistic kinematics. 
The most important result obtained was that the relativistic kinematics 
improves the relative position of the positive- and 
negative-parity nucleon excited states \cite{DESP,CARL} as compared to 
the results obtained by means of the same potential in 
a nonrelativistic description \cite{BRAC}. However, 
the complete inversion that is observed experimentally 
is not achieved.

The so-called level ordering problem in the nucleon spectrum
can be easily understood in the pure harmonic limit.
The $N^*(1440)$ $J^P=1/2^+$ belongs to the $[56,0^+]$ 
$SU(6)_{FS} \times O(3)$
irreducible representation and it appears in the $N=2$ band,
while the $N^*(1535)$ $J^P=1/2^-$ belongs to the
$[70,1^-]$ appearing in the $N=1$ band. As a consequence,
the $N^*(1440)$ has $2 \hbar \omega$ energy excitation while the 
$N^*(1535)$ has only $1 \hbar \omega$ energy excitation, 
opposite to the order observed experimentally. Theoretically,
this situation has been cured by means of appropriate phenomenological
interactions as has been the case of anharmonic terms \cite{ISGU},
scalar three-body forces \cite{BERT} or pseudoscalar interactions
\cite{GAR2,GLOZ}.

The understanding of the few body problem in terms of 
interactions between quarks requires a 
more complicated dynamics than the one provided
by the OGE. In particular, as dictated by the breaking
of chiral symmetry a Goldstone boson exchange interaction
between quarks will appear. In an effective chiral symmetric
description of the baryon spectrum to lowest order a 
pseudoscalar one-pion-exchange (OPE)
interaction between quarks is present. 
Such interaction is known to be very important,
responsible among other things for the
long-range part of the nucleon-nucleon potential \cite{VAL1}.
Along with the pseudoscalar boson chiral symmetry requires 
an accompanying scalar meson field to complete the chiral multiplet.
This also contributes an effective attractive interaction spin and 
flavor-independent, the $\sigma$-meson exchange. 
Such a model provides with a reasonable
description of the two and three non-strange baryon systems
\cite{VAL1,GAR4}.

We have already studied the baryon spectrum by means of a dynamical model 
that considers Goldstone boson exchanges between quarks 
besides the standard OGE. In Ref. \cite{GAR2} we 
presented calculations for the baryon spectrum performed 
in momentum space but with a nonrelativistic kinematics.
In this paper we want to focus our attention on the effect
of the relativistic kinematics for the description of the low-energy
baryon spectrum. In particular, the possible lowering of the first radial
nucleon excitation with respect to the $L=1$ state \cite{DESP,CARL}.
Besides, we want to study the enhancement of this effect due
to the flavor-spin structure of the
pseudoscalar boson-exchange interaction. For this purpose we will
make use of a quite standard quark-quark interaction composed of
a confining potential, a one-gluon exchange, and 
Goldstone-boson exchanges.

Let us briefly resume the most important aspects of the solution
of the Schr\"odinger equation with relativistic kinematics.
If one replaces in the Schr\"odinger equation the nonrelativistic 
kinetic energy operator
by the corresponding relativistic expression 
it becomes

\begin{equation}
|\psi> = G_0(W_0)[V_1+V_2+V_3]|\psi>,
\label{eq37}
\end{equation}
where if one assumes that the three
particles are in the c.m. system, i.e.,

\begin{equation}
\vec k_1+\vec k_2+\vec k_3=0,
\label{eq41}
\end{equation}
then $W_0$ is the invariant mass of the system and

\begin{equation}
G_0(W_0) = {1\over W_0-
\omega_1(k_1) - \omega_2(k_2) - \omega_3(k_3) },
\label{eq38}
\end{equation}
with

\begin{equation}
\omega_i(k_i) = \sqrt{m_i^2+k_i^2}.
\label{eq36}
\end{equation}

Making the Faddeev decomposition

\begin{equation}
|\psi> = |\phi_1>+|\phi_2>+|\phi_3>,
\label{eq8}
\end{equation}
one obtains the Faddeev equations

\begin{equation}
|\phi_i> = G_0(W_0)t_i(W_0)[|\phi_j>+|\phi_k>],
\label{eq39}
\end{equation}
with

\begin{equation}
t_i(W_0)=V_i+V_i G_0(W_0) t_i(W_0).
\label{eq40}
\end{equation}
Our basis states are $|\vec p_i\vec q_i>$ 
where $\vec p_i$ is the 
relative momentum of the pair $jk$ measured in the c.m. frame of the 
pair (that is, the frame in which particle $j$ has momentum $\vec p_i$ and
particle $k$ has momentum $-\vec p_i$) and $\vec q_i = -\vec k_i$ is the
relative momentum between the pair $jk$ and particle $i$ measured in the 
three-body c.m. frame (that is, the frame in which the pair $jk$ has 
total momentum $\vec q_i$ and particle $i$ has momentum $-\vec q_i$).
The total energy of the three particles 
$\omega_1(k_1) + \omega_2(k_2) + \omega_3(k_3)$ 
can be written in terms of
the relative momenta $\vec p_i$ and $\vec q_i$ as
\begin{equation}
\omega_1(k_1) + \omega_2(k_2) + \omega_3(k_3)
= W_i(p_i q_i) + \omega_i(q_i),
\label{eq42}
\end{equation}
where 
\begin{equation}
W_i(p_i q_i) = \sqrt{\omega^2(p_i)+q_i^2},
\label{eq43}
\end{equation}
and
\begin{equation}
\omega(p_i) = \sqrt{m_j^2+p_i^2}+\sqrt{m_k^2+p_i^2}\equiv\omega_j(p_i)
+\omega_k(p_i).
\label{eq44}
\end{equation}
The invariant volume element for three particles satisfying Eq.
(\ref{eq41}) can be written in terms of the corresponding volume element
for the relative momenta as
\begin{equation}
{d\vec k_1\over 2\omega_1(k_1)}{d\vec k_2\over 2\omega_2(k_2)}{d\vec k_3
\over 2\omega_3(k_3)}\,\delta(\vec k_1+\vec k_2+\vec k_3)=
{\omega(p_i)\over 8W_i(p_i q_i)\omega_i(q_i)\omega_j(p_i)
\omega_k(p_i)}\,d\vec p_i d\vec q_i.
\label{eq46}
\end{equation}
Therefore, if the single-particle states are normalized invariantly on the 
mass shell, i.e.,
\begin{equation}
<\vec k_i|\vec k_i^{\,\prime}>=2\omega_i(k_i)
\delta(\vec k_i - \vec k_i^{\,\prime}),
\label{eq47}
\end{equation}
then the basis states $|\vec p_i\vec q_i>$ are normalized as
\begin{equation}
<\vec p_i \vec q_i|\vec p_i^{\,\prime}\vec q_i^{\,\prime}> =
{8W_i(p_i q_i)\omega_i(q_i)\omega_j(p_i)
\omega_k(p_i)\over
\omega(p_i)}
\delta(\vec p_i - \vec p_i^{\,\prime})
\delta(\vec q_i - \vec q_i^{\,\prime}),
\label{eq48}
\end{equation}
and satisfy the completeness relation
\begin{equation}
1=\int 
{\omega(p_i)\over 8W_i(p_i q_i)\omega_i(q_i)\omega_j(p_i)
\omega_k(p_i)}d\vec p_i d\vec q_i\,|\vec p_i \vec q_i><\vec p_i\vec q_i|.
\label{eq49}
\end{equation}

Introducing complete sets of basis states
into the operator equations (\ref{eq39}) and (\ref{eq40}) one obtains 
the integral equations 

\begin{eqnarray}
<\vec p_i\vec q_i|\phi_i> & = & G_0(W_0;p_iq_i)\sum_{j\ne i}\int 
{\omega(p_i^\prime)\over 8W_i(p_i^\prime q_i^\prime)
\omega_i(q_i^\prime)
\omega_j(p_i^\prime)
\omega_k(p_i^\prime)}
d{\vec p_i}^{\,\prime}
d{\vec q_i}^{\,\prime} 
\nonumber \\  & & \times 
{\omega(p_j)\over 8W_j(p_j q_j)\omega_j(q_j)\omega_k(p_j)
\omega_i(p_j)}
d\vec p_j d\vec q_j <\vec p_i\vec q_i|t_i(W_0)|
{\vec p_i}^{\,\prime}{\vec q_i}^{\,\prime}>
\nonumber \\  & & \times 
<{\vec p_i}^{\,\prime}{\vec q_i}^{\,\prime}|\vec p_j\vec q_j><\vec p_j\vec q_j|
\phi_j>, 
\label{eq50}
\end{eqnarray}
and
\begin{eqnarray}
<\vec p_i\vec q_i|t_i(W_0)|
{\vec p_i}^{\,\prime}\vec q_i^{\,\prime}> & = &
<\vec p_i\vec q_i|V_i|
{\vec p_i}^{\,\prime}\vec q_i^{\,\prime}>+\int 
{\omega(p_i^{\prime\prime})
\over 8W_i(p_i^{\prime\prime}q_i^{\prime\prime})\omega_i(q_i^{\prime\prime})
\omega_j(p_i^{\prime\prime})
\omega_k(p_i^{\prime\prime})}
d{\vec p_i}^{\,\prime\prime}
d\vec q_i^{\,\prime\prime}
\nonumber \\  & & \times
<\vec p_i\vec q_i|V_i|
{\vec p_i}^{\,\prime\prime}\vec q_i^{\,\prime\prime}>
G_0(W_0;p_i^{\prime\prime}q_i^{\prime\prime}) 
<{\vec p_i}^{\,\prime\prime}\vec q_i^{\,\prime\prime}|t_i(W_0)|
{\vec p_i}^{\,\prime}\vec q_i^{\,\prime}>,
\label{equ51}
\end{eqnarray}
where the matrix elements of the potential 
are given by \cite{GAR1}
\begin{eqnarray}
<\vec p_i\vec q_i|V_i|\vec p_i^{\,\prime}\vec q_i^{\,\prime}> & = &
8\omega_i(q_i)\left[{W_i(p_iq_i)\omega_j(p_i)\omega_k(p_i)
W_i(p_i^\prime q_i)\omega_j(p_i^\prime)\omega_k(p_i^\prime)\over
\omega(p_i)\omega(p_i^\prime)}\right]^{1/2}
\nonumber \\  & & \times 
\delta(\vec q_i -{\vec q_i}^{\,\prime})V_i
(\vec p_i,
{\vec p_i}^{\,\prime}),
\label{equ52}
\end{eqnarray}
with $V_i(\vec p_i,
{\vec p_i}^{\,\prime})$ the usual Fourier transform of the potential
\begin{equation}
V_i(\vec p_i,
{\vec p_i}^{\,\prime})={1\over (2\pi)^3}\int d\vec r\,
e^{i\vec p_i\cdot\vec r}V(r)e^{-i\vec p_i^{\,\prime}\cdot\vec r}.
\label{equ10}
\end{equation}
As shown in Ref. \cite{GAR1} these equations can be
partial-wave projected into the basis 
$|p_iq_i;\ell_i\lambda_i>_L$ where $\ell_i$ and $\lambda_i$ are the orbital
angular momentum quantum numbers corresponding to the momenta 
$\vec p_i$ and $\vec q_i$, respectively, and $L$ is the total orbital
angular momentum. Since the two-body interactions that we
consider do not have tensor nor spin-orbit terms the spin is treated 
nonrelativistically by means of Racah coefficients in the same way as
the isospin \cite{GAR1}. We solved the integral equations including
all the configurations with 
$\ell_i$ and $\lambda_i$ up to 5.

As has been already mentioned we will use a quark-quark 
potential of the type
\begin{equation}
V_{qq} (\vec{r}_{ij}) = V_{CON} (\vec{r}_{ij}) + 
V_{OGE} (\vec{r}_{ij})
+ V_{OPE} (\vec{r}_{ij})  + V_{OSE} (\vec{r}_{ij}) \, ,
\label{eq1}
\end{equation}

\noindent
where $\vec{r}_{ij}$ is the interquark distance $(\vec{r}_{ij} = 
\vec{r}_{i} - \vec{r}_{j})$.
$V_{CON}$ is the confining potential taken to be linear,
\begin{equation}
V_{CON} ({\vec r}_{ij}) =
- a_c \, {\vec
\lambda}_i \cdot {\vec \lambda}_j
\,r_{ij} \, .
\label{eq2}
\end{equation}

\noindent
$V_{OGE}$ is the one-gluon-exchange potential

\begin{equation}
V_{OGE} ({\vec r}_{ij}) =
{1 \over 4} \, \alpha_s \, {\vec
\lambda}_i \cdot {\vec \lambda}_j
\Biggl \lbrace {1 \over r_{ij}} -
{1 \over {4 \, m^2_q}} \, \biggl [ 1 + {2 \over 3}
{\vec \sigma}_i \cdot {\vec
\sigma}_j \biggr ] \,\,
{{e^{-r_{ij}/r_0}} \over
{r_0^2 \,\,r_{ij}}}
 \Biggr \rbrace \, ,
\label{eq3}
\end{equation}

\noindent
$\alpha_s$ is the effective quark-quark-gluon coupling constant, the
$\lambda 's$ are the $SU (3)$ color matrices, and the $\sigma ' s$
stand for the spin Pauli matrices.
$V_{OPE}$ and $V_{OSE}$ are the pseudoscalar and scalar Goldstone-boson
exchange potentials, given respectively by:

\begin{equation}
V_{\rm OPE} ({\vec r}_{ij})  =  {1 \over 3}
\, \alpha_{ch} {\Lambda^2  \over \Lambda^2 -
m_\pi^2} \, m_\pi \,  \left[ \,
Y (m_\pi \, r_{ij}) - { \Lambda^3
\over m_{\pi}^3} \, Y (\Lambda \,
r_{ij}) \right] {\vec \sigma}_i \cdot
{\vec \sigma}_j 
{\vec \tau}_i \cdot {\vec \tau}_j \, ,
\label{eq4}
\end{equation}

\begin{equation}
V_{\rm OSE} ({\vec r}_{ij}) = - \alpha_{ch} \,
{4 \, m_q^2 \over m_{\pi}^2}
{\Lambda^2 \over \Lambda^2 - m_{\sigma}^2}
\, m_{\sigma} \, \left[
Y (m_{\sigma} \, r_{ij})-
{\Lambda \over {m_{\sigma}}} \,
Y (\Lambda \, r_{ij}) \right] \, ,
\label{eq5}
\end{equation}

\noindent
where $m_\pi$ ($m_\sigma$) is the pion (sigma) mass,
$\alpha_{ch}$ is the chiral coupling constant related to the
$\pi NN$ coupling constant through $\alpha_{ch} =
\left( 3 \over 5 \right)^2
{ g_{\pi NN}^2 \over {4 \pi}} { m_{\pi}^2
\over {4 m_N^2}}$, $\Lambda$ is a cutoff
parameter, and $Y(x)$ is the standard Yukawa function
$Y (x) =  e^{-x}/x$

The one-gluon-exchange potential shown in Eq. (\ref{eq3})
contains a smeared-out
delta function with smearing parameter $r_0$. As it has been shown in
Refs. \cite{GAR1,GAR2} this smeared delta function has a very important
effect on the baryon spectrum. For example, in the case of the 
nonrelativistic theory \cite{GAR2}, if $r_0 \le 0.1$ fm the 
mass of the nucleon collapses to very large negative values. On the 
other hand, in the case of the relativistic
theory the collapsing of the nucleon mass occurs for 
$r_0 \le 0.45$ fm due to the fact that in this case 
the three-body propagator falls down more slowly 
in momentum space \cite{GAR1}.

The behavior of the solution of the three-body equations in
the presence of a delta function will have also important 
consequences when one includes the one-pion-exchange interaction. 
As can be seen in Eq. (\ref{eq4}) the one-pion-exchange potential 
consists of a Yukawa interaction with a range equal to 
the inverse of the pion mass plus a smeared-out delta function with 
smearing parameter $1/\Lambda$. Therefore, also in this case the 
mass of the nucleon will collapse if $1/\Lambda \le 0.1$ fm in 
the case of the nonrelativistic theory and if $1/\Lambda \le 0.45$  
fm in the case of the relativistic one. Thus, the cutoff parameter 
$\Lambda$ must be smaller than $\sim$ 10 fm$^{-1}$ in the case 
of the nonrelativistic theory and $\sim$ 2.2
fm$^{-1}$ in the case of the relativistic one. This is due to
the fact that the nonrelativistic propagator falls down as
$1/k^2$ while the relativistic one does it as $1/k$.
This is, for example, the reason why the semirelativistic solutions of
Ref. \cite{FURU} collapse when a scalar and an one-gluon-exchange
potential are added to the pseudoscalar interaction. They include a strong
one-gluon exchange attraction, $1/\Lambda_g \sim 0.20$ fm,
over the strong attractive interaction
already provided by the one-pion exchange, $1/\Lambda_\pi \sim 0.29$ fm.
This makes evident the risk of replacing
the nonrelativistic kinetic energy by the relativistic one
when the details of the short-range part of the potential are not well known.

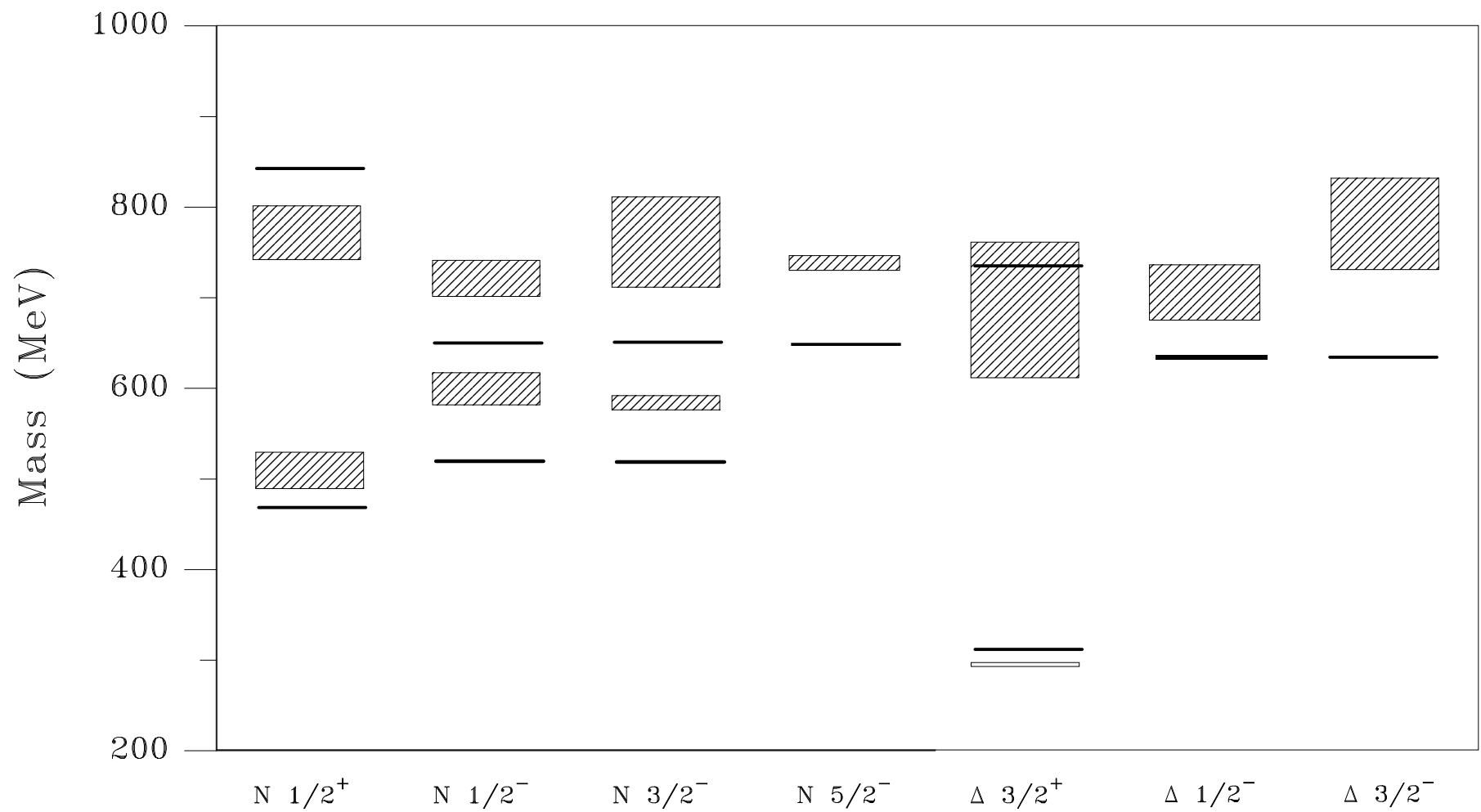
\begin{figure}
\caption{Nonstrange baryon spectrum for the interaction given
in Eq. (\ref{eq1}) with $V_{OSE}=0$ and the set of parameters 
of Table \ref{tab1}.
Masses are given relative to the nucleon mass.}
\label{fig1}
\end{figure}

Let us first try to illustrate how the combined effect of
relativistic kinematics and the one-pion-exchange interaction 
gives rise to the inversion between the first positive- and negative-parity 
nucleon excited states. For this purpose we consider the
quark-quark interaction given in Eq. (\ref{eq1}) with
$V_{OSE}=0$ and the set of parameters of Table \ref{tab1}. 
The obtained spectrum is shown in Fig. \ref{fig1}. As can be seen
the positive parity $N^*(1440)$, Roper resonance, lies below the 
negative parity $N^*(1535)$ state, while the remaining states of 
the spectrum are correctly reproduced. The same effect has been 
observed in a nonrelativistic study (see Fig. 4 of Ref. \cite{GAR2})
at the price of loosing the description of the remaining states
of the spectrum. It is only when relativistic kinematics is
introduced, that the inversion between positive- and 
negative-parity excitations of the nucleon is achieved with an
overall good description of the rest of the spectrum. The relativistic
kinematics alone decreases a lot the excitation energy of the 
Roper resonance with respect to the first excited state of negative
parity as compared to a nonrelativistic description \cite{CARL}. The 
Roper resonance is particularly sensitive to the form of the kinetic
energy \cite{CARL}. Once this energy difference 
has been decreased, the one-pion
exchange is able to produce the desired
inversion without destroying the description of the
rest of the spectrum.

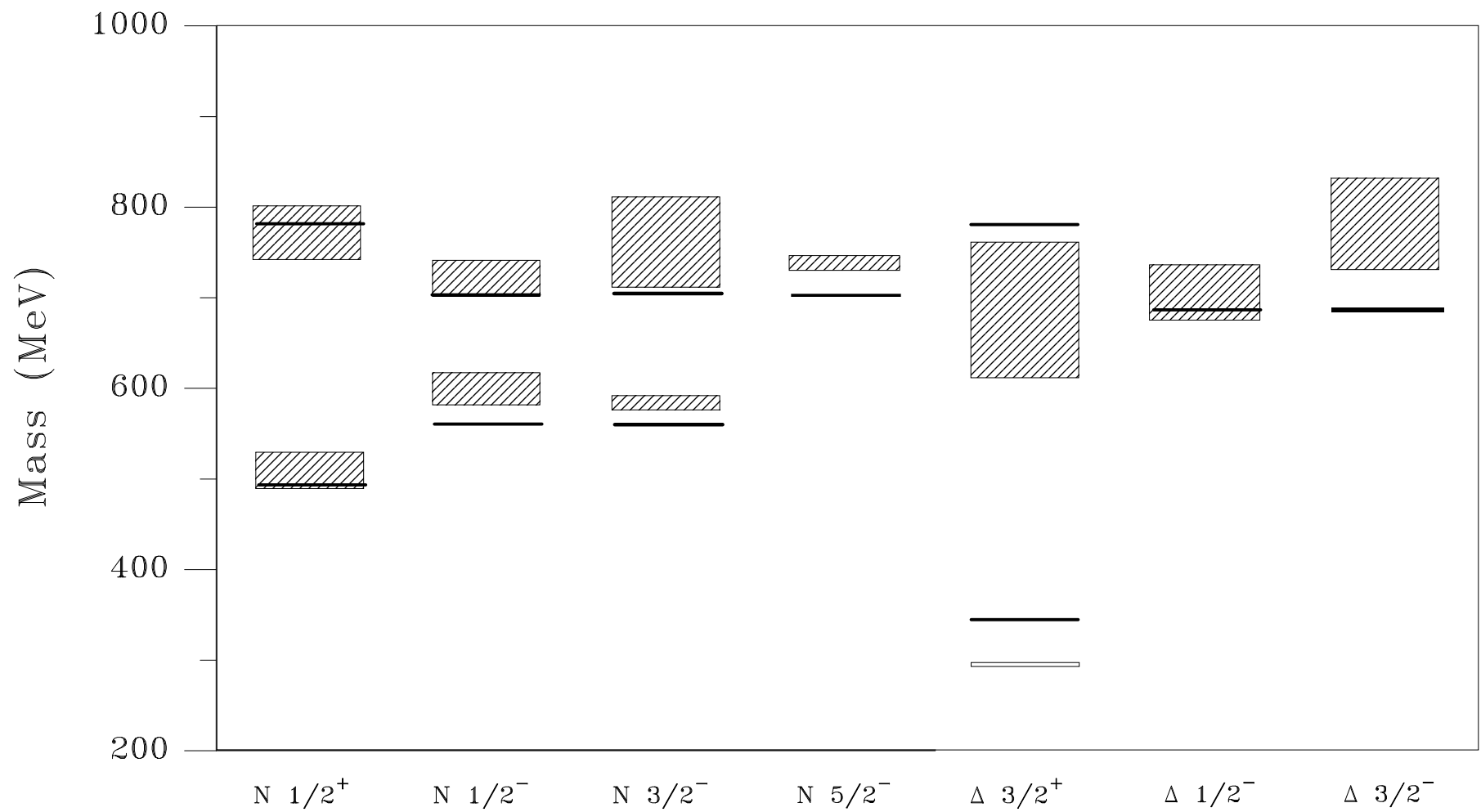
\begin{figure}
\caption{Nonstrange baryon spectrum for the interaction given
in Eq. (\ref{eq1}) and the set of parameters of Table \ref{tab1}.
Masses are given relative to the nucleon mass.}
\label{fig2}
\end{figure}

Chiral symmetry requires also the presence of an scalar Goldstone
boson exchange. Although in a perturbative description its effect
would be minor, we have calculated the spectrum obtained when one takes also 
into account the scalar interaction given by 
Eq. (\ref{eq5}) with the same set of parameters shown in 
Table \ref{tab1}. The results are presented in Fig. \ref{fig2}.
As one can see, the agreement with the experimental data is much better 
than in the previous case. It reduces the splitting between the
$J^P=1/2^+$ states $N^*(1440)$ and $N^*(1710)$, and increases
the distance between the ground and first excited
states of negative parity. On the other hand, the $\Delta$(1232)
comes out somewhat higher although we have not made any effort 
to fine tune the parameters of the model, our main objective
being the effect on the relative position of the positive and
negative parity nucleon states.

The reverse of the ordering achieved by means of the pion exchange
is due to a similar mechanism than the one proposed in Ref. \cite{BERT}
by means of a scalar three-body force. Such a force is very 
short-ranged and in the limit of zero range it acts only for 
states whose wave functions do not cancel at the origin. 
It therefore affects the nucleon and its radial
excitations, while producing essentially no effect for states with
mixed symmetry (negative parity nucleon excited states). Thus, 
this force, if it is chosen attractive, explains
why the Roper resonance is lower than the negative 
parity excited states in the nucleon sector.

In the case of the chiral pseudoscalar interaction, its
$(\vec{\sigma} \cdot \vec{\sigma}) (\vec{\tau} \cdot \vec{\tau})$ 
structure gives attraction for symmetric spin-isospin pairs and
repulsion for antisymmetric ones (a quite distinctive
feature since the color-magnetic part of the OGE
gives similar contributions in both cases).
This lowers the position of the first nucleon radial excitation
[$N^*(1440)$] with a completely symmetric spin-isospin
structure with regard to the
first negative parity state [$N^*(1535)$], with a mixed symmetry
wave function in spin-isospin space. Due to the decreasing of the
excitation energy of the Roper resonance induced by the relativistic
kinematics, the pseudoscalar force is able to solve
the discrepancy between usual
two-body potential models (the predicted relative energy positions
of the Roper and the first negative parity
appear inverted) and experiment. 

A final comment regarding the one-gluon-exchange interaction
is in order. As has been explained this potential alone is not able 
to reverse the order of the Roper resonance and the first $N(1/2^-)$
state \cite{GAR1}. However, we did not find any limit to the strength of this
potential for a correct description of the spectrum once relativistic
kinematics and the one-pion exchange are introduced. 
The parameters we have used for the one-gluon exchange
are the standard values used to study the $NN$ interaction 
$\alpha_s \sim 0.4 - 0.5$. One could imagine that a better fit 
to the baryon spectrum could be obtained by decreasing the strength of the
one-gluon exchange, what could provide with a bigger inversion
as explained in the previous paragraph.
However, the effect obtained 
is exactly the opposite one. In Fig. \ref{fig2} the mass difference between
the Roper resonance and the first $N(1/2^-)$ state is $-70$ MeV. If we 
recalculate the same spectrum using $\alpha_s =0.4$ instead of
$\alpha_s = 0.5$, there is a general decreasing of the mass of
all the states, the $\Delta(3/2^+)$ appears at 270 MeV and the Roper
resonance appears at 465 MeV, that could be cured increasing the slope 
of the confining potential. However, the mass difference between the 
Roper resonance and the first $N(1/2^-)$ state is reduced to $-31$ MeV.
Therefore, once the relativistic kinematics 
and the one-pion exchange are considered
the effect of the one-gluon exchange cannot be estimated qualitatively,
because it generates repulsion on both states but the strength depends
on the wave function, and therefore on the other interactions considered.
This observation reinforces the notion that the one-gluon-exchange
force is necessary for the description of the $NN$ interaction \cite{NAKA}.

As a summary, we have shown that the relative position of the 
positive and negative parity low-energy states is determined
basically by the relativistic kinematics together with the
one-pion-exchange interaction. A great improvement is obtained
by adding the chiral partner of the pion, the sigma meson.
We did not find any constrain to the strength of the one-gluon
exchange potential. Finally, let us stress that the replacement
of the nonrelativistic by the relativistic kinematics for short-range
attractive interactions has to be done with a careful study of the
short-range part of the potential in the relativistic approach.

\acknowledgements
We are specially grateful to Dr. B. Desplanques for calling our attention
on the problem raised in this work and giving us a lot of technical and
practical information. We also thank to Dr. P. Gonz\'alez
for enlightening
discussion. This work has been partially funded by COFAA-IPN 
(M\'exico), by Ministerio de Ciencia y Tecnolog{\'{\i}}a 
under Contract No. BFM2001-3563 and by Junta de Castilla y Le\'{o}n 
under Contract No. SA-109/01.

\begin{table}
\caption{Quark model parameters for the calculation of Figs. \ref{fig1}
and \ref{fig2}.}
\label{tab1}

\begin{tabular}{cccc}
 & $m_q ( {\rm MeV})$                          &  313       & \\
 & $\alpha_s$                                  &  0.50      & \\
 & $a_c ( {\rm MeV} \cdot {\rm fm}^{-1})$      &  110.0    & \\
 & $\alpha_{ch}$                               &  0.0269    & \\
 & $r_0 ({\rm fm})$                            &  0.74       & \\
 & $m_\sigma ({\rm fm}^{-1})$                  &  3.42      & \\
 & $m_\pi ({\rm fm}^{-1})$                     &  0.7       & \\
 & $\Lambda ({\rm fm}^{-1})$                   &  2.0       & \\
\end{tabular}
\end{table}

\end{document}